# Magnetic Resonance Elastography and Portal Hypertension: Influence of the Portal Venous Flow on the Liver Stiffness

Simon CHATELIN[a,1], Raoul POP[b,c], Céline GIRAUDEAU[b], Khalid AMBARKI[d], Ning JIN[e], François SEVERAC[a,f], Elodie BRETON[a] and Jonathan VAPPOU[a]

[a] *ICube, CNRS UMR 7357, University of Strasbourg, IHU, Strasbourg, France*
[b] *IHU-Strasbourg, Institute for Image-Guided Surgery, Strasbourg, France*
[c] *University Hospital of Strasbourg, Strasbourg, France*
[d] *Siemens Healthcare SAS, Saint Denis, France*
[e] *Siemens Medical Solutions, Chicago, IL, USA*
[f] *Laboratory of Biostatistics and Medical Computer Science, School of Medicine, University Hospital of Strasbourg, Strasbourg, France*

**Abstract.** The invasive measurement of the hepatic venous pressure gradient is still considered as the reference method to assess the severity of portal hypertension. Even though previous studies have shown that the liver stiffness measured by elastography could predict portal hypertension in patients with chronic liver disease, the mechanisms behind remain today poorly understood. The main reason is that the liver stiffness is not specific to portal hypertension and is also influenced by concomitant pathologies, such as cirrhosis. Portal hypertension is also source of a vascular incidence, with a substantial diversion of portal venous blood to the systemic circulation, bypassing the liver. This study focuses on this vascular effect of portal hypertension. We propose to generate and control the portal venous flow (to isolate the modifications in the portal venous flow as single effect of portal hypertension) in an anesthetized pig and then to quantify its implications on liver stiffness by an original combination of MRE and 4D-Flow Magnetic Resonance Imaging (MRI). A catheter balloon is progressively inflated in the portal vein and the peak flow, peak velocity magnitude and liver stiffness are quantified in a 1.5T MRI scanner (AREA, Siemens Healthcare, Erlangen, Germany). A strong correlation is observed between the portal peak velocity magnitude, the portal peak flow or the liver stiffness and the portal vein intraluminal obstruction. Moreover, the comparison of mechanical and flow parameters highlights a correlation with the possibility of identifying linear relationships. These results give preliminary indications about how liver stiffness can be affected by portal venous flow and, by extension, by hypertension.

## 1. Introduction

In healthy volunteers, the portal vein represents about 87% of the total liver inflow [1], as illustrated in the figure 1. As a major consequence of chronic liver disease, portal hypertension is responsible for significant morbidity and mortality [2]. To the best of our knowledge and despite its invasiveness, the measurement of the hepatic venous pressure gradient is still considered as the reference method to assess the severity of portal hypertension [3]–[5]. The hepatic venous pressure gradient is an important indicator for the decision for portal vein embolization and thus to reduce the risk of liver failure after hepatic tumor resection. Portal hypertension is not only characterized by an increase of hepatic venous pressure gradient but is also source of vascular effects, with a substantial diversion of portal venous blood to the systemic circulation, bypassing the liver [6]. Cirrhosis and non-cirrhotic portal fibrosis are important causes for portal hypertension. During the last decade, elastography has been proposed as a noninvasive alternative to the measurement of hepatic venous pressure gradient for detecting the severity of portal hypertension with clinical usefulness [7]–[10]. Especially in case of severe degree, portal hypertension has been shown to be related on the amount of portal blood inflow [11], [12]. Previous studies have shown that the mechanical properties (at first the liver stiffness) accurately predicts portal hypertension in patients with chronic liver disease, with comparable performance to the hepatic venous pressure gradient [8], [10].

[1] Corresponding author. E-mail address: schatelin@unistra.fr (S. Chatelin)   - *Diffusion non restreinte*

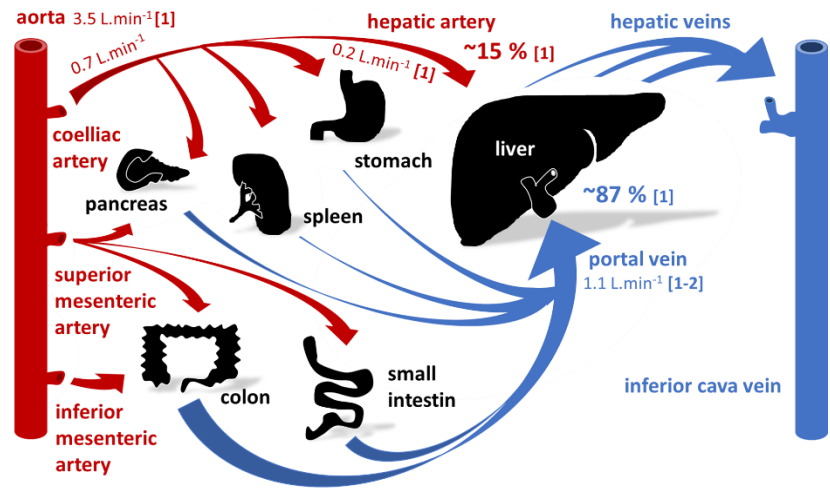

**Figure 1.** Schematic illustration of the portal and hepatic vascular system. (sources for the human healthy flow values: [1: [1]] [2: [13]])

However, little is known about the relationship between venous pressure and the liver mechanical properties - mainly limited to systematic reviews and meta-analysis [12] - and remain still today poorly understood, despite an obvious correlation [13]. The main reason is that most of the commercial and experimental elastography imaging systems estimate liver stiffness as a mechanical parameter, which is correlated but not specific to portal hypertension and at the same time influenced by several pathologies, such as liver fibrosis and cirrhosis [14], [15].

This study aims at generating and isolating the modifications in the portal venous flow as a single effect of portal hypertension to highlight their influence on liver stiffness. Instead of studying correlations between liver stiffness and portal hypertension in heathy and pathological cohorts, we propose to control the portal venous flow (using an inflatable balloon catheter) and to quantify its effects on both the portal flow characteristics and the liver stiffness by the original combination of 4D-Flow Magnetic Resonance Imaging (4D-Flow MRI) and Magnetic Resonance Elastography (MRE).

## 2. Materials and Methods

To isolate the modifications in the portal venous flow as single effect of portal hypertension, the portal venous flow is controlled, quantified and the effects on liver stiffness are investigated by an original combination of MRE and 4D-Flow MRI.

The tests are performed on 1 female pig weighting 27 kg. Anesthesia is achieved using intravenous Zoletil (*20cc, Virbac, Carros, France*) and then maintained using uniform ventilation (except during MRI sequences requiring breath holding) with 2% isoflurane (*Forene, AbbVie, North Chicago, IL, USA*). The experiments are performed in agreement with the European Community Council Directive of 22. September 2010 (*010/63/UE*) and the French local ethics committee (*protocol 38.2018.01.103 - APAFIS #14092-2018031513247711 v1*).

The portal venous flow is controlled by partial or complete occlusion of the portal vein using a MR-compatible balloon catheter (*XXLTM Vascular, Boston Scientific, Marlborough, MA, USA*), with 9 mm in diameter and 40 mm in length. Using fluoroscopic guidance, the balloon catheter is advanced in the main trunk of the portal vein. Balloon inflation is first performed under fluoroscopy in order to determine the inflation volume necessary for varying degrees of portal occlusion. A 1:1 contrast (Visipaque 270, GE Healthcare, Buckinghamshire, United Kingdom) saline solution is used for balloon inflation.

The catheter balloon is progressively inflated in the portal vein and the flow and mechanical parameters are quantified *in vivo* in a 1.5T interventional MRI scanner (MAGNETOM *Aera, Siemens Healthcare, Erlangen, Germany*). The protocol is divided in five successive blocs (steps), each one of them corresponding to a specific inflation of the balloon in the portal vein: 0%, 50%, 80%, 0% and 100% of intraluminal sectional obstruction of the portal vein, successively. Each bloc (step) is composed of 4 different MRI sequences acquired in a 1.5T MAGNETOM Aera MRI interventional scanner (Siemens Healthcare, Erlangen, Germany) and corresponding to:

(1) The validation of the position of the field-of-view (FOV) on low resolution anatomical coronal images (T1 VIBE sequence, hold breathing, TR/TE = 4.72/2.18 ms, FOV 243 mm x 300 mm, matrix 156 x 256, 72 slices, slice thickness 2 mm).

(2) The measurement of the liver stiffness (MRE gradient recalled echo (GRE) sequence with motion sensitizing gradients, hold breathing, TR/TE = 50/23.75 ms, FOV 239 mm x 300 mm, matrix 204 x 256, 3 slices, slice thickness 5 mm). The stiffness maps are obtained using a local estimation algorithm in a 3D region of interest over the 3 coronal slices.

(3) The portal venous flow characterization by the quantitative 3D investigation over time of the peak flow and the peak velocity magnitude in the portal vein. The acquisitions are performed using the prototype 4D-Flow pulse sequence [16] (Siemens Healthcare, Erlangen, Germany) (free breathing and respiratory gating, TR/TE = 51.68/3.77 ms, FOV 284 mm x 414 mm, matrix 106 x 192, 20 slices, slice thickness 2.5 mm, 11 time frames, encoding velocity 50 cm.s$^{-1}$ for all directions, TT = 20 to 440 by 52.5 steps for 9 temporal frames). The post-treatment and the flow parameters quantification are performed using the prototype "*4DFlow v2.4*" WIP software (Siemens Healthcare, Erlangen, Germany) with phase anti-aliasing and motion tracking filtering before flow quantifications.

(4) The precise measurement of the balloon and portal vein dimensions using high resolution anatomical coronal images (prototype sequence: SPIRAL VIBE sequence, free breathing, TR/TE = 4.57/0.05 ms, FOV 300 mm x 300 mm, matrix 288 x 288, 176 slices, slice thickness 1 mm).

The respiratory synchronization of the MRI sequences is ensured using MR-compatible electrocardiogram. The angiography image with 80% inflated balloon (A), 4D-Flow image (B), MRE stiffness map (C) and SPIRAL VIBE anatomical image (D) are shown in figure 2.

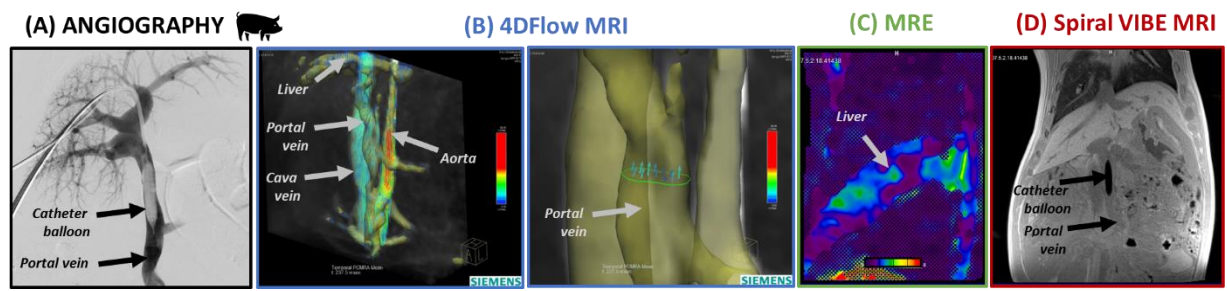

*Figure 2.* A catheter balloon is inserted in the portal vein under angiography (A) and then successively inflated at 0%, 50%, 80%, 0% and 100% intraluminal sectional occlusion of the vein. The portal venous flow (B) and liver mechanical (C) parameters are successively measured for each occlusion state using 4D-Flow MRI and MRE imaging. Spiral VIBE MRI sequence images and measures the precise dimensions of the balloon (D).

## 3. Results and discussions

From the SPIRAL VIBE MRI images, the successive intraluminal sectional vein occlusions are investigated by the measurement of both the inflated catheter balloon and the portal vein diameter. With successive values of 0.6 mm, 7.9 mm, 5.8 mm, 0.6 mm, 9.1 mm and 8.7 mm, 8.9 mm, 7.9 mm, 8.7 mm, 9.1 mm for the balloon and portal vein diameter, respectively, the theoretical successive sectional obstruction values of 0%, 80%, 50%, 0% and 100% are adjusted to 0.4%, 79.8%, 52.5%, 0.4% and 100.0% real successive occlusion values.

The insertion of the catheter balloon is thought to modify the vascular flow and velocity, first of all in the portal vein. As illustrated in figure 3(A), a significant correlation appears between the portal peak flow and the inflation state of the balloon. Similarly, the portal peak velocity magnitude is correlated to the occlusion of the portal vein. Otherwise, this trend is not observed in the cava vein (figure 3(B)). In the cava vein, the velocity and flow parameters decrease slowly and constantly over the experiments, independently of the inflation of the catheter balloon. The experimental conditions, and above all to the anesthesia, could explain this slow and regular flow decrease in the cava vein [17], [18]. These observations validate the specificity of our results to the portal vascular system.

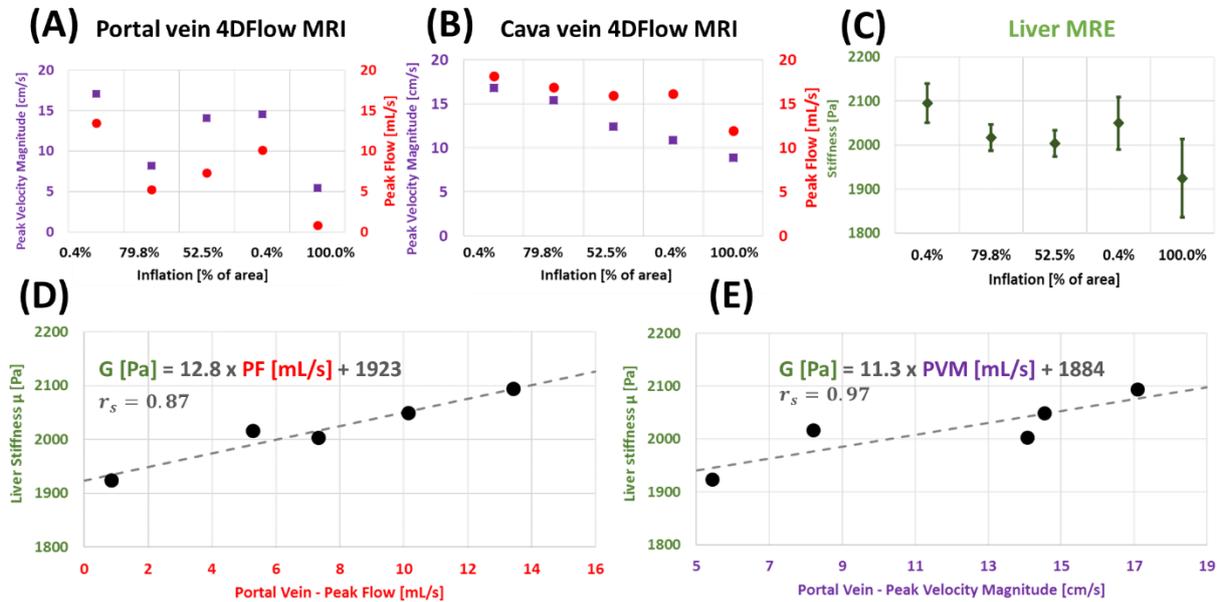

**Figure 3.** The portal (A) and cava vein (B) flow parameters are quantified using 4D-Flow imaging over portal vein intraluminal occlusion (peak velocity magnitude and peak flow in purple squares and red circles, respectively). The liver stiffness (C) is measured using MRE over portal vein intraluminal occlusion (in green diamond shape). The investigation of the liver stiffness as a function of portal peak flow (D) and portal peak velocity magnitude (E) indicates a linear relationship.

From the MRE measurements, a similar trend as for portal peak flow or portal peak velocity magnitude over portal occlusion state can be observed for liver stiffness (figure 3(C)). This result illustrates a direct influence of the portal flow obstruction on the liver mechanical properties measured by elastography.

The liver stiffness is observed as a function of the portal venous peak flow and portal peak velocity magnitude, as reported in the figures 3 (D) and (E), respectively. The individual correlations between the liver stiffness and flow parameters are performed using non-parametric Spearman's $r_s$ rank correlation coefficients to investigate potential monotonic relationships. A significant correlation is observed, with the possibility of identifying linear relationships, as attested by the correlation coefficient close of 0.87 and 0.97 for peak flow and peak velocity magnitude, respectively. These linear relationships are identified for this single animal as:

$$\mu[Pa] = 12.8 \times PF[mL/s] + 1923 \quad \text{and} \quad \mu[Pa] = 11.3 \times PVM[mL/s] + 1884$$

$\mu$, PF and PVM are the liver stiffness (from MRE measurements), the portal peak flow and the portal peak velocity magnitude (from 4D-Flow MRI measurements), respectively.

## 4. Conclusions

The results and conclusions of this preliminary study are limited by the observation of a case-specific single subject and could be strengthened by increasing the cohort to a complete prospective study. Nevertheless, these preliminary results are already thought to give indications about how liver stiffness can be affected by portal inflow alterations and so by portal hypertension. They contribute to give information about the effect, which would affect the elastography measurements of the liver in clinical practice. Otherwise, this study provides unprecedented *in vivo* information for the mechanical modeling of the liver including mechanical properties and vascular flows.

**Acknowledgements**

The authors thank Thomas Benkert (*Siemens Healthcare GmbH, Germany*) for provided the prototype SPIRAL VIBE sequence.

**Fundings**

This work was supported by French state funds managed by the ANR (*Agence Nationale de la Recherche*) within the Investissements d'Avenir program for the IHU Strasbourg (*IHU Strasbourg, Institut Hospitalo-Universitaire, Institute of Image Guided Surgery, ANR-10-IAHU-02*).